# Thermally-Induced Structural Evolution of Silicon- and Oxygen-Containing Hydrogenated Amorphous Carbon: A Combined Spectroscopic and Molecular Dynamics Simulation Investigation

Filippo Mangolini[1], James Hilbert[2], J. Brandon McClimon[3], Jennifer R. Lukes[2], and Robert W. Carpick[2,*]

[1] Institute of Functional Surfaces, School of Mechanical Engineering, University of Leeds, Woodhouse Ln., Leeds, West Yorkshire LS2 9JT, UK
[2] Department of Mechanical Engineering and Applied Mechanics, University of Pennsylvania, Philadelphia, Pennsylvania 19104, USA
[3] Department of Materials Science and Engineering, University of Pennsylvania, Philadelphia, Pennsylvania 19104, USA

\* To whom correspondence should be addressed:
Professor Robert W. Carpick
Department of Mechanical Engineering and Applied Mechanics,
University of Pennsylvania
229 Towne Bldg., 220 S. 33$^{rd}$ Street
19104-6315 Philadelphia, PA, USA

Phone: +1-215-898-4608
Fax: +1-215-573-6334
carpick@seas.upenn.edu

**Abstract (max. 600 characters, including spaces. Currently: 585 characters)**

The thermally-induced structural evolution of silicon- and oxygen-containing hydrogenated amorphous carbon (a-C:H:Si:O) was investigated by X-ray photoelectron and absorption spectroscopy, as well as molecular dynamics (MD) simulations. The spectroscopic results indicate that the introduction of Si and O in hydrogenated amorphous carbon (a-C:H) increases the activation energy for the conversion of $sp^3$- to $sp^2$-bonded C. MD simulations indicate that the higher thermal stability of a-C:H:Si:O compared to a-C:H derives from the lower fraction of strained C-C $sp^3$ bonds in a-C:H:Si:O.

**Text (max. 3750 words, including figures. Currently: 3851 words)**

Amorphous carbon (a-C) materials are in widespread use due to many attractive properties including high strength, low friction, and ease of applying them as thin, conformal coatings on a wide range of substrates [1]. Doping and alloying these materials are effective methods to achieve multifunctionality as well as to boost existing properties or introduce new ones [1-3]. Among the several doped a-C materials that have been synthesized over the last decades, silicon- and oxygen-containing hydrogenated amorphous carbon (a-C:H:Si:O) is an important class of multicomponent materials for several applications, since incorporating silicon and oxygen reduces their residual stress [4, 5], while not significantly affecting the mechanical properties [5]. In addition, a-C:H:Si:O films present other interesting properties, such as enhanced thermal stability [6, 7] and good tribological behavior (*i.e.*, low friction and wear) across a broader range of conditions and environments compared to hydrogenated amorphous carbon (a-C:H) [5, 8, 9]. The superior thermal stability of a-C:H:Si:O compared to a-C:H was suggested to arise from the fourfold coordination of silicon, which stabilizes the carbon atoms in the $sp^3$ hybridization state (thus inhibiting their conversion into $sp^2$-bonded carbon at elevated temperatures) [10-13]. Yet the influence of silicon and oxygen on the energetics of the structural evolution of a-C:H:Si:O in response to energetic inputs has not been investigated. The challenges associated with the characterization and quantification of the bonding configuration of carbon in carbon-based materials make the study of the structure of a-C films experimentally difficult [1, 14]. Atomistic simulations can yield insights into the evolution of these materials in response to energetic inputs since they allow structural processes to be identified [15, 16]. Here, we present the results from a spectroscopic investigation of the thermally-induced structural evolution of a-C:H:Si:O, with accompanying molecular dynamics (MD) simulation to help elucidate the mechanisms of enhanced thermal stability.

a-C:H:Si:O films (Sulzer-Metco Inc., Amherst, USA) were deposited on silicon wafers using a proprietary plasma-enhanced chemical vapor deposition process [4, 5, 9, 17-22]. To investigate *in situ* the thermally-induced structural evolution of a-C:H:Si:O, heating experiments were performed inside X-ray photoelectron spectroscopy (XPS) and near edge X-ray absorption fine structure (NEXAFS) spectroscopy chambers. To develop a better understanding of the atomic-scale processes occurring in a-C:H:Si:O at elevated temperatures, MD simulations were performed using the ReaxFF potential [23-27]. A detailed description of the procedures for performing the experiments, acquiring/processing the XPS/NEXAFS data, and carrying out MD simulations is reported in the Supporting Information (SI), and further details about the MD simulations will be reported elsewhere [28].

The C 1s signal of unannealed a-C:H:Si:O (Figure 1a and 1b) was fit with six components at 283.23±0.05 (silicon bound to carbon as in silicon carbide [29, 30]), 284.46±0.05 eV (a-C:H:Si:O, which includes carbon bound to carbon and hydrogen as in a-C:H [31] and carbon bound to silicon as in organosiloxanes [32-34]), 285.00±0.05 eV (aliphatic contamination – due to air exposure [29, 32]), and 286.44±0.06 eV, 287.82±0.07 eV, and 289.16±0.09 eV (carbon bound to oxygen – contamination due to air exposure [32, 35]). The Si 2p spectrum of unannealed a-C:H:Si:O (Figure 1c) turned out to contain four doublets, whose maxima in their $2p_{3/2}$ components were found at 100.8±0.1 eV, 101.6±0.1 eV, 102.5±0.1 eV, and 103.6±0.1 eV. These values are typical for Si(I) (as well as Si bound to C as in silicon carbide), Si(II), Si(III), and Si(IV), respectively [30, 36-38]. On the basis of XPS data, the composition of as-deposited a-C:H:Si:O was calculated: [C] = 83.3±0.6 at.%; [O] = 7.6±0.4 at.%; [Si] = 9.1±0.5 at.% (note: the composition calculated by XPS does not include hydrogen. The composition of a-C:H:Si:O measured by Rutherford backscattering spectrometry (RBS) and hydrogen forward scattering (HFS) spectrometry (Evans Analytical Group, USA) was: [C] = 57±3 at.%; [O] = 3±1 at.%; [Si] = 6±1 at.%; [H] = 34±3 at.% [22, 39]). The composition of the as-deposited material calculated by XPS, together with the outcomes of characterization of a-C:H:Si:O by RBS, HFS spectrometry, and X-ray reflectivity (reported in the SI), were used as an input parameter in the MD simulations. The simulated, unannnealed structure of a-C:H:Si:O is composed of a single, dense, fully amorphous phase with no evidence of porosity (Figure S.1). The extremely low percentage of carbon-oxygen bonds in the MD-simulated structure of a-C:H:Si:O (<2%) supports the assumption made in XPS data processing that carbon atoms bound to oxygen are in the contamination layer (*i.e.*, not in a-C:H:Si:O) due to the sample exposure to air.

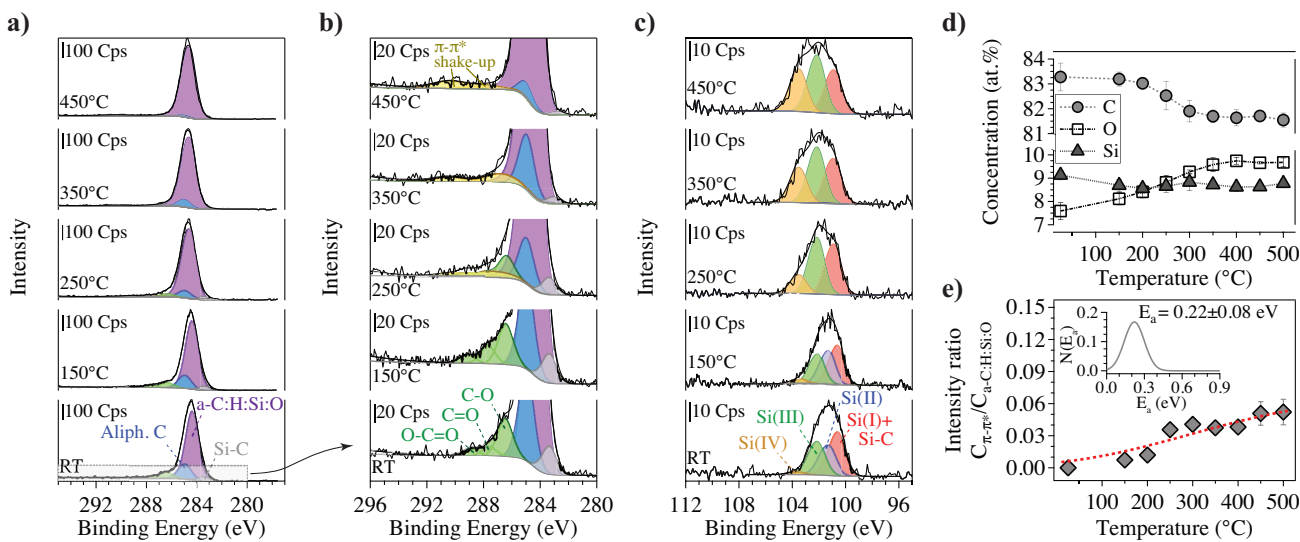

**Figure 1.** (a)-(c) High-resolution XPS spectra ((a)-(b) C 1s, (c) Si 2p) of a-C:H:Si:O acquired before and after annealing under high vacuum conditions. Colored lines are curve fits. (b) is a zoomed view from (a). The C 1s spectra indicate the progressive desorption of the carbonaceous contamination layer on as-deposited a-C:H:Si:O upon annealing together with clustering and ordering of $sp^2$-bonded carbon. The Si 2p signals indicate a slight increase in the relative fraction of Si in higher oxidation states with annealing temperature; (d) composition of a-C:H:Si:O calculated from XPS data as a function of annealing temperature; (e) intensity ratio between the $\pi-\pi^*$ shake-up satellites and the a-C:H:Si:O synthetic peak *vs.* annealing temperature, indicating thermally-induced clustering and ordering of $sp^2$-hybridized carbon sites. The red line is a fit with a model (described in the SI) based on an Arrhenius equation including a Gaussian distribution of activation energies. The computed activation energy range for clustering and ordering of $sp^2$-hybridized carbon (inset) agrees well with the activation energy for the same process in a-C:H [31].

Upon heating at temperatures above 200°C, the composition of a-C:H:Si:O slightly changed, namely, [C] decreased, while [O] increased (Figure 1d). This could be due to the reaction of a-C:H:Si:O with adsorbed water or residual oxygen/water in the XPS chamber. Progressive changes in the XPS spectra were also observed upon annealing. The curve synthesis of the Si 2p signal indicated a slight increase in the relative fractions of silicon in higher oxidation states (+3 and +4) with annealing temperature (Figure 1c). This finding is supported by MD simulations: upon annealing the MD-simulated structure of a-C:H:Si:O for 60 ps at 275°C, 525°C, and 825°C, the progressive increase in the fraction of silicon-oxygen bonds together with the decrease in the number of silicon-hydrogen and silicon-carbon bonds (Figure 2a and S.2a) strongly suggest an increase in the silicon oxidation state. The decrease in the number of silicon-hydrogen bonds observed in MD simulations is in agreement with *ex situ* infrared spectroscopic measurements (Figure S.3), which indicated the disappearance of the characteristic Si-H stretching vibration peak upon annealing at 450°C under high vacuum conditions (Figure S.3).

To properly calibrate the intensity of the XPS peaks contributing to the C 1s spectra for the annealed samples, the total area of the C 1s peaks assigned to carbon bound to oxygen was set equal to the total area (corrected for the sensitivity factors) of the peaks for oxygen bound to carbon obtained from fitting the O 1s spectrum (not shown), once the contribution for oxygen bound to silicon (calculated from fitting the Si 2p signal) was subtracted from the O 1s signal. Upon annealing, the contamination layer on as-deposited a-C:H:Si:O progressively desorbed, as suggested by the decrease of the intensities of the peaks assigned to aliphatic carbon and carbon bound to oxygen (Figure 1a and 1b). In addition, the lineshape of the C 1s peak at 284.5±0.1 eV changed in the spectra of the annealed samples: while at room temperature this peak was symmetric, a good fit to the experimental data after annealing could only be obtained by introducing a tail on the high binding-energy side of the synthetic curve. The C 1s spectra of a-C:H:Si:O annealed at temperatures above 200°C also exhibited the appearance of two new components at 287.4±0.3 eV and 290.4±0.2 eV (Figure 1b), whose intensity increased with annealing temperature. The appearance of these peaks, which were assigned to the characteristic π-π* shake-up satellites in ordered $sp^2$-hybridized carbon [40, 41], together with the increased lineshape asymmetry of the C 1s peak, suggest a progressive clustering and ordering of the $sp^2$ carbon sites upon annealing together with a reduction of the dihedral angle between π electrons [31, 40, 41]. MD simulations support the formation of clusters of $sp^2$-bonded carbon atoms upon annealing through the progressive increase in the relative fraction of $sp^2$-hybridized neighbors around each $sp^2$-bonded carbon atom (Figure 2b and S.2b).

It has been postulated that $sp^2$-hybridized carbon sites act like defects in the $sp^3$-hybridized carbon matrix: upon providing thermal energy, they can diffuse and form larger clusters [31, 42]. For example, for a hydrogen-free tetrahedral amorphous carbon (ta-C) film, Ferrari *et al.* calculated the activation energy (0.28 eV) for the clustering and ordering of $sp^2$ carbon on the basis of Raman spectroscopic data [42]. The authors of the present work provided evidence that the clustering of $sp^2$ carbon sites in an a-C:H film is a thermally-activated process with a distribution of activation energies with mean value equal to 0.18 eV and standard deviation equal to 0.05 eV [31]. To compute the activation energy for the clustering and ordering of $sp^2$ carbon in a-C:H:Si:O, the ratio of the intensity of the π-π* shake-up satellites to that of the a-C:H:Si:O synthetic peak was plotted as a function of temperature (Figure 1e). The experimental data was fit with a model (described in the SI) based on an Arrhenius equation including a Gaussian distribution of activation energies, and using the area, mean and standard deviation of the Gaussian distribution as fitting parameters (correlation coefficient: 0.85). The computed activation energy range (0.22±0.08 eV) for the clustering of $sp^2$ carbon in a-C:H:Si:O agrees well with the activation energy range for the same process in a-C:H [31], thus indicating that the introduction of silicon and oxygen in a-C:H at levels of, respectively, 6±1 at.% and 3±1 at.% to form this a-C:H:Si:O film has no significant effect on the energetics of the clustering and ordering of $sp^2$ carbon.

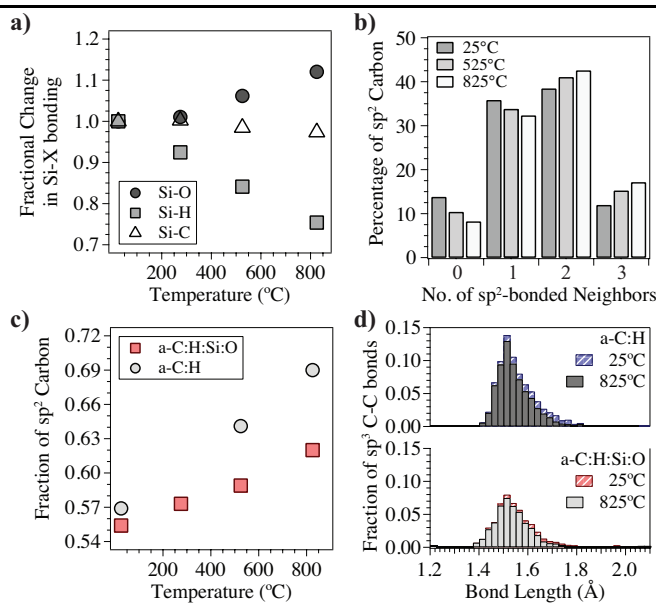

**Figure 2.** MD simulations of the thermally-induced structural evolution of a-C:H:Si:O: (a) fractional change in Si-X bonds (where X = O, C, or H) *vs.* annealing temperature. The absolute values of the frequency of these bonds are provided in the SI (Figure S.2a). Upon annealing, the fraction of Si-O bonds progressively increases, while the number of Si-H and Si-C bonds decreases; (b) percentage of $sp^2$-bonded carbon with different numbers of neighboring $sp^2$-bonded carbon atoms before and after annealing. The relative change in the number of $sp^2$-hybridized neighbors around each $sp^2$-bonded carbon atom upon annealing is provided in the SI (Figure S.2b). The results indicate the formation of clusters of $sp^2$-hybridized carbon upon annealing; (c) fraction of $sp^2$-hybridized carbon *vs.* annealing temperature for a-C:H:Si:O and a-C:H, indicating the thermally-induced $sp^3$-to-$sp^2$ rehybridization of carbon atoms and the higher thermal stability of a-C:H:Si:O compared to a-C:H; and (d) fraction of $sp^3$-hybridized C-C bonds *vs.* bond length for a-C:H:Si:O and a-C:H before and after annealing at 825ºC. The higher thermal stability of a-C:H:Si:O compared to a-C:H derives from the lower fraction of strained C-C $sp^3$ bonds in a-C:H:Si:O.

To quantitatively investigate *in situ* the changes in the bonding configuration of carbon in a-C:H:Si:O as a function of temperature, NEXAFS spectroscopic measurements were performed in ultrahigh vacuum (UHV) conditions. NEXAFS spectra obtained at the carbon 1s edge (corrected for the contribution of the contamination layer following the method outlined in Ref. [43]) on the as-deposited and *in situ* annealed a-C:H:Si:O samples are displayed in Figure 3a and 3b. The NEXAFS spectrum of as-deposited a-C:H:Si:O exhibited a feature at 285.0±0.1 eV, which is due to the C 1s→π* transition for disordered carbon-carbon bonds [44, 45]. The intensity of this peak directly correlates with the fraction of $sp^2$-bonded carbon in the near-surface region [43, 45-48]. A broad hump between 288 and 310 eV, which is due to the C 1s→σ* transition for disordered carbon-carbon σ bonds [44, 45], characterized the NEXAFS spectrum of as-deposited a-C:H:Si:O. The presence of carbon-hydrogen and carbon-silicon bonds in a-C:H:Si:O also resulted in the detection of distinct features at 287.0±0.1 eV (C 1s→σ* transition for C-H bonds [43, 44, 47]) and 288.9±0.1 eV (C 1s→σ* transition for C-Si bonds [49]).

The C 1s NEXAFS spectra of a-C:H:Si:O annealed *in situ* at different temperatures are also displayed in Figure 3a and 3b. Upon annealing, the intensity of the peak at 285.0±0.1 eV increased, suggesting an increase in the fraction of $sp^2$-bonded carbon in the near-surface region. At the same time, the absorption feature at 288.9±0.1 eV decreased in intensity with the annealing temperature, thus indicating the breakage of carbon-silicon bonds. The feature at 287.0±0.1 eV also changed at elevated temperatures: its intensity first increased upon annealing at 150°C, which might be due to hydrogen diffusion to the near-surface region upon low temperature annealing, and then progressively decreased, which suggests the scission of carbon-hydrogen bonds upon annealing. These spectral variations could be highlighted by computing the difference in NEXAFS spectra between subsequent annealing steps (Figure S.4). Calculating these difference spectra also allowed the detection of a progressive increase in spectral intensity on the high photon energy side (at ~285.5 eV) of the peak assigned to the C 1s→π* transition for disordered carbon-carbon bonds, which indicates an increase in the degree of ordering of the $sp^2$ bond (in agreement with XPS results) [44, 45]. Atomistic details from MD simulations support these thermally-induced structural transformations occurring in a-C:H:Si:O: upon annealing the MD-simulated structure of a-C:H:Si:O for 60 ps at 275°C, 525°C, and 825°C, a progressive increase in the fraction of $sp^2$-bonded carbon was observed (Figure 2c), whereas the number of silicon-carbon bonds decreases (Figure 2a and S.2a).

The fraction of $sp^2$-hybridized carbon, which was calculated from the NEXAFS spectra following the method reported in Ref. [48], is displayed as a function of annealing temperature in Figure 3c. In the case of ta-C, the transformation of $sp^3$- to $sp^2$-hybridized carbon was modeled by Sullivan *et al.* as a series of first order chemical reactions [50]. Because of bond length and angle disorder in ta-C, Sullivan *et al.* postulated the presence of a distribution of activation energies for $sp^3$-to-$sp^2$ conversion of carbon hybridization (found between 1 eV and 3 eV). Grierson *et al.*, who investigated the thermally-induced evolution of the structure of ta-C by NEXAFS spectroscopy, applied the model of Sullivan *et al.* to determine an activation energy range for $sp^3$-to-$sp^2$ conversion of carbon hybridization [46]. The energy range calculated by Grierson *et al.* (3.5±0.9 eV) agreed with the activation energy for bulk $sp^3$-to-$sp^2$ conversion of carbon hybridization in ta-C calculated on the basis of Raman spectroscopic measurements (3.3. eV) [42]. The variation of $sp^2$ fractions *vs.* temperature calculated on the basis of the data presented above was fit with the model of Sullivan *et al.* assuming the distribution of activation energies for $sp^3$-to-$sp^2$ conversion to be Gaussian, and using the mean and standard deviation of the Gaussian distribution as fitting parameters (a detailed description of the model is reported in the SI). The broad distributions of bond lengths and angles computed from MD simulations of the unannealed structure of a-C:H:Si:O (Figure S.5) support the assumption made by Sullivan *et al.* concerning the presence of a distribution of activation energies for $sp^3$-to-$sp^2$ conversion of carbon hybridization due to bond length/angle disorder. The experimental data could be well fit with the model proposed by Sullivan *et al.* (correlation coefficient: 0.99) and yielded an activation energy distribution for $sp^3$-to-$sp^2$ conversion of 3.0±1.1 eV. This energy range is lower than the one calculated for the $sp^3$-to-$sp^2$ conversion of carbon-carbon bonds in ta-C [42, 46], but higher than the activation energy necessary for directly converting $sp^3$- into $sp^2$-hybridized carbon in a-C:H (when the thermally-induced structural transformation of a-C:H is fit with the model of Sullivan *et al.* [31]). This finding suggests that introducing silicon and oxygen in a-C:H affects the energetics of the $sp^3$-to-$sp^2$ conversion of carbon hybridization and enhances the thermal stability in high vacuum compared to undoped a-C:H. MD simulations supported this finding: for a-C:H:Si:O there is substantially less relative change in the fraction of $sp^3$-bonded carbon with annealing compared to a-C:H (with similar fraction of $sp^3$-bonded carbon and hydrogen content) (Figure 2c). Atomistic insights into the origin of the higher thermal stability of a-C:H:Si:O compared to a-C:H (with similar fraction of $sp^3$-bonded carbon and hydrogen content) could be gained by MD simulations. Annealing a-C:H at 825°C for 60 ps demonstrates that strained C-C bonds are more likely to break (Figure 2d) and lead to an increase in the fraction of $sp^2$ carbon. The introduction of silicon in a-C:H results in a narrower distribution of C-C bonds with less strained bonds due to the Si-C greater average bond length compared to the C-C (185 pm compared to 154 pm [51]) (Figure 2d and S.6), which allows for the structure to accommodate a greater degree of structural disorder without as many highly strained bonds, while increasing the thermal stability.

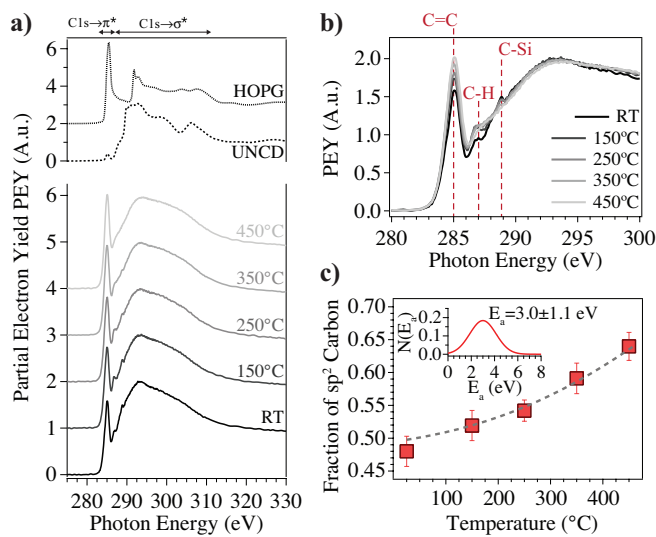

**Figure 3.** (a) C 1s NEXAFS spectra of a-C:H:Si:O acquired before and after annealing under UHV conditions. The C 1s spectra of reference samples (*i.e.*, HOPG and UNCD) are also displayed (dashed lines). Spectra are pre- and post-edge normalized as well as offset for clarity; (b) zoomed view of the absorption edge region of the C 1s NEXAFS spectra. Spectra displayed without any offset to allow for comparisons. The results indicate that upon annealing there is an increase in the fraction of $sp^2$-bonded carbon together with the breakage of C-Si bonds. Additionally, the evolution of the C-H spectral feature indicates hydrogen diffusion to the near-surface region upon annealing at 150°C and scission of C-H bonds at higher temperatures; (c) fraction of $sp^2$-hybridized carbon *vs.* annealing temperature calculated from NEXAFS spectra. The experimental data were fit with the model of Sullivan *et al.* [50] (black dashed line). Inset: the corresponding distribution of activation energies for $sp^3$-to-$sp^2$ conversion. The computed activation energy is lower than the one for the $sp^3$-to-$sp^2$ conversion of C-C bonds in ta-C [42, 46], but higher than the one for converting $sp^3$- into $sp^2$-bonded C in a-C:H [31]. This suggests that introducing silicon and oxygen in a-C:H enhances the thermal stability in high vacuum compared to a-C:H.

In summary, the thermally-induced structural evolution of a-C:H:Si:O was investigated by *in situ* XPS/NEXAFS spectroscopic experiments and MD simulations. Upon high vacuum annealing, two thermally activated processes with an assumed Gaussian distribution of activation energies with mean value $E$ and standard deviation $\sigma$ occur in a-C:H:Si:O: (a) ordering and clustering of $sp^2$-hybridized carbon ($E$=0.22 eV; $\sigma$=0.08 eV); and (b) conversion of $sp^3$- to $sp^2$-hybridized carbon ($E$=3.0 eV; $\sigma$=1.1 eV). Atomistic details from MD simulations provide a qualitative validation of these processes occurring in a-C:H:Si:O at elevated temperatures (a quantification of the energetics of the thermally-induced processes could not be performed at this time because of the different time- and length-scales between experiments and simulations). The comparison of the outcomes of the present work with previous results [31] indicated that introducing silicon and oxygen in the a-C:H network enhances the thermal stability under high vacuum conditions. Atomistic details from MD simulations suggested that the higher thermal stability of a-C:H:Si:O compared to a-C:H (with similar fraction of $sp^3$-bonded carbon and hydrogen content) derives from the significantly lower fraction of strained C-C $sp^3$ bonds in a-C:H:Si:O compared to a-C:H, which are more likely to break at elevated temperatures.

This material is based upon work supported by the Advanced Storage Technology Consortium ASTC (grant 2011-012) and the National Science Foundation under Grant No. DMR-1107642. F.M. acknowledges support from the Marie Curie International Outgoing Fellowship for Career Development within the 7$^{th}$ European Community Framework Programme under contract no. PIOF-GA-2012-328776 and the Marie Skłodowska-Curie Individual Fellowship within the European Union's Horizon 2020 Program under Contract No. 706289. The authors would like to thanks Dr. C. Jaye, Dr. D.A. Fischer, Dr. P. Albrecht, and Dr. D.R. Mullins for their kind assistance with the NEXAFS measurements at the National Synchrotron Light Source. Use of the National Synchrotron Light Source, Brookhaven National Laboratory, was supported by the US Department of Energy, Office of Science, Office of Basic Energy Sciences, under Contract No. DE-AC02-98CH10886. We are grateful to Dr. F. Rose (previously at HGST, a Western Digital Company) for the X-ray reflectivity measurements.